\documentclass[twocolumn]{aastex631}
\usepackage[version=4]{mhchem}
\usepackage[T1]{fontenc}

\begin{document}

\title{CAMEMBERT: A Mini-Neptunes GCM Intercomparison, Protocol Version 1.0.\\
A CUISINES Model Intercomparison Project.}

\author[0000-0002-4997-0847]{Duncan A. Christie}
\affiliation{Physics and Astronomy, College of Engineering, Mathematics and Physical Sciences, University of Exeter, Exeter EX4 4QL, UK}

\author[0000-0002-3052-7116]{Elspeth K. H. Lee}
\affiliation{Center for Space and Habitability, University of Bern, Gesellschaftsstrasse 6, CH-3012 Bern, Switzerland}

\author[0000-0001-5271-0635]{Hamish Innes}
\affiliation{Atmospheric, Oceanic and Planetary Physics, University of Oxford, UK}

\author[0000-0002-8012-3400]{Pascal A. Noti}
\affiliation{Center for Space and Habitability, University of Bern, Gesellschaftsstrasse 6, CH-3012 Bern, Switzerland}

\author[0000-0003-0977-6545]{Benjamin Charnay}
\affiliation{LESIA, Observatoire de Paris, Université PSL, CNRS, Sorbonne Université, Université de Paris, 5 place Jules Janssen, 92195 Meudon, France}

\author[0000-0002-5967-9631]{Thomas J. Fauchez}
\affiliation{NASA Goddard Space Flight Center,
8800 Greenbelt Road,
Greenbelt, MD 20771, USA}
\affiliation{Integrated Space Science and Technology Institute, Department of Physics, American University, Washington DC}
\affiliation{NASA GSFC Sellers Exoplanet Environments Collaboration}

\author[0000-0001-6707-4563]{Nathan J. Mayne}
\affiliation{Physics and Astronomy, College of Engineering, Mathematics and Physical Sciences, University of Exeter, Exeter EX4 4QL, UK}

\author[0000-0001-9423-8121]{Russell Deitrick}
\affiliation{School of Earth and Ocean Sciences, University of Victoria, Victoria, British Columbia, Canada}

\author[0000-0001-7758-4110]{Feng Ding}
\affiliation{School of Engineering and Applied Sciences, Harvard University, Cambridge, MA 02138, USA}

\author[0000-0002-4649-1568]{Jennifer J. Greco}
\affiliation{Department of Physics and Astronomy, University of Toledo, 2801 W. Bancroft Street, Toledo, OH 43606, USA}

\author[0000-0002-6893-522X]{Mark Hammond}
\affiliation{Atmospheric, Oceanic and Planetary Physics, University of Oxford, UK}

\author[0000-0003-0217-3880]{Isaac Malsky}
\affiliation{Department of Astronomy, University of Michigan, Ann Arbor, MI 48109, USA}

\author[0000-0002-8119-3355]{Avi Mandell}
\affiliation{NASA Goddard Space Flight Center, 8800 Greenbelt Road, Greenbelt, MD 20771, USA}

\author[0000-0003-3963-9672]{Emily Rauscher}
\affiliation{Department of Astronomy, University of Michigan, Ann Arbor, MI 48109, USA}

\author[0000-0001-8206-2165]{Michael T. Roman}
\affiliation{School of Physics and Astronomy, University of Leicester, University Road, Leicester LE1 7RH, UK}
\affiliation{Facultad de Ingeniería y Ciencias, Universidad Adolfo Ibáñez, Av. Diagonal las Torres 2640, Peñalolén, Santiago, Chile}

\author[0000-0001-8832-5288]{Denis E. Sergeev}
\affiliation{Physics and Astronomy, College of Engineering, Mathematics and Physical Sciences, University of Exeter, Exeter EX4 4QL, UK}

\author[0000-0002-6673-2007]{Linda Sohl}
\affiliation{Center for Climate Systems Research, Columbia University, New York, NY, USA}
\affiliation{NASA Goddard Institute for Space Studies, 2880 Broadway, New York, NY 10025, USA}

\author[0000-0001-8342-1895]{Maria E. Steinrueck}
\affiliation{Max Planck Institute for Astronomy, D-69117 Heidelberg, Germany}

\author[0000-0003-2260-9856]{Martin Turbet}
\affiliation{Laboratoire de M\'et\'eorologie Dynamique/IPSL, CNRS, Sorbonne Universit\'e, \'Ecole Normale Sup\'erieure, PSL Research University, \'Ecole Polytechnique, 75005 Paris, France}

\author[0000-0002-7188-1648]{Eric T. Wolf}
\affiliation{Laboratory for Atmospheric and Space Physics, Department of Atmospheric and Oceanic Sciences, University of Colorado, Boulder,
Boulder, CO}
\affiliation{NASA GSFC Sellers Exoplanet Environments Collaboration, Greenbelt, MD}
\affiliation{NExSS Virtual Planetary Laboratory, Seattle, WA}

\author[0000-0002-9705-0535]{Maria Zamyatina}
\affiliation{Physics and Astronomy, College of Engineering, Mathematics and Physical Sciences, University of Exeter, Exeter EX4 4QL, UK}

\author[0000-0001-9355-3752]{Ludmila Carone}
\affiliation{Centre for Exoplanet Science, School of Physics \& Astronomy, University of St Andrews, North Haugh, St Andrews KY169SS, UK}



\begin{abstract}
With an increased focus on the observing and modelling of mini-Neptunes, there comes a need to better understand the tools we use to model their atmospheres.  In this paper, we present the protocol for the CAMEMBERT (Comparing Atmospheric Models of Extrasolar Mini-neptunes Building and Envisioning Retrievals and Transits) project, an intercomparison of general circulation models (GCMs) used by the exoplanetary science community to simulate the atmospheres of mini-Neptunes.   We focus on two targets well studied both observationally and theoretically with planned JWST Cycle 1 observations:  the warm GJ~1214b and the cooler K2-18b.  For each target, we consider a temperature-forced case, a clear sky dual-grey radiative transfer case, and a clear sky multi band radiative transfer case, covering a range of complexities and configurations where we know differences exist between GCMs in the literature.   This paper presents all the details necessary to participate in the intercomparison, with the intention of presenting the results in future papers.   Currently, there are eight GCMs participating ({\sc ExoCAM}, {\sc Exo-FMS}, FMS PCM, {\sc Generic PCM}, {\sc MITgcm}, RM-GCM, THOR, and the UM), and membership in the project remains open.  Those interested in participating are invited to contact the authors.
\end{abstract}



\section{Introduction}

Super-Earths and mini-Neptunes represent a demarcation, albeit a nebulous one, between the giant planets with their thick atmospheres dominated by hydrogen and helium and the terrestrial planets with thinner secondary atmospheres \citep{lopez2014}.  The planets that have retained their hydrogen-dominated atmospheres are believed to have undergone runaway accretion during their formation in order to accumulate a thick atmosphere \citep{pollack1996,lee2014} but have also been able to retain some or all of that atmosphere in the presence of irradiative evaporation \citep{owen2012}.

Simulating these planets with general circulation models (GCMs) presents a unique set of challenges not necessarily seen in the Earth sciences community.  The primitive equations, which assume hydrostatic balance, a thin atmosphere, and a constant gravitational acceleration with height, may not yield accurate results for cases where the thickness of the modelled atmosphere becomes significant relative to the radius of the planet, limiting their applicability to planets with thick atmospheres (see \citealt{tokano2013}, \citealt{tort2015}, and \citealt{Mayne2019} for a discussions related to Venus and Titan, Earth, and mini-Neptunes, respectively). It has also been argued that simulations of mini-Neptunes specifically may have extremely long convergence times, potentially of 50,000 Earth days of model time or more \citep{wang2020}, which raises questions about the accuracy of simulations of only a few thousand days, as are common in the exoplanetary modelling community \citep[e.g.,][]{charnay2015,Mayne2017,Mayne2019}.  These issues motivate a better understanding of the tools we use to model these planets.

While intercomparison studies have been somewhat common in the Earth sciences community (see e.g. \citealt{cmip6,rfmip2016,hiresmip2016,dcmip2016}), it has not been until recently that intercomparisons have been done with a focus on exoplanetary targets.  Although not an intercomparison of multiple GCMs, \citet{heng2011} performed a comparison of the spectral and finite-difference dynamical cores in the Geophysical Fluid Dynamics Laboratory (GFDL) Princeton Flexible Modelling System (FMS) using the hot Jupiter HD209458~b as a test case. The first true intercomparison of GCMs used in exoplanetary science, \citet{polichtchouk2014}, looked at highly idealised configurations -- a steady state jet, a baroclinic wave, and diabatic forcing -- for five GCMs with the intention of better understanding their respective dynamical cores.  Increasing the complexity, \citet{yang2019} compared GCMs in the context of terrestrial planets, specifically focusing on the cases of an Earth-like planet orbiting a G-star and a tidally-locked planet around an M-star.  More recently, the TRAPPIST-1 Habitable Atmosphere Intercomparison  \citep[THAI;][]{thaiprotocol} compared GCM models of TRAPPIST-1e, investigating the dynamics \citep{thai1,thai2} and the synthetic observations \citep{thai3} resulting from the simulations.  In the final paper, they propose a ``GCM uncertainty error bar'' of $\sim 50\%$ when interpreting transmission spectra with the uncertainty explained mostly by the cloud differences found between GCMs. The ability to provide this form of context to synthetic observations highlights the value of projects like THAI.

Based on the success of the THAI project, we propose here the CAMEMBERT (Comparing Atmospheric Models of Extrasolar Mini-neptunes Building and Envisioning Retrievals and Transits) intercomparison of GCMs modelling mini-Neptunes under the umbrella of the CUISINES (Climates Using Interactive Suites of Intercomparisons Nested for Exoplanet Studies) framework for intercomparisons for exoplanets \citep{cuisinesreport}.  The broad objectives of CUISINES are twofold. First, it provides a meta-framework to quantify, and potentially mitigate, differences between exoplanet model outputs. Second, it aims to assess how these output differences affect the synthetic observations that are used to predict the detectability of atmospheric constituents and to interpret data from ground and space telescopes. With the increased focus on mini-Neptunes with the launch of TESS \citep[e.g.,][]{trifonov2019,lacedelli2021,burt2021}, CHEOPS \citep[e.g.,][]{bonfanti2021,leleu2021}, JWST \citep[e.g.,][]{greenejwst2017,beanjwst2021,hujwst2021} and in anticipation of increased efforts to model these planets, we believe the timing is appropriate for an intercomparison of GCMs modelling mini-Neptunes to provide a foundational understanding of how our models behave and how our model choices may impact the interpretation of observations.

In this first paper, we outline the protocol for the CAMEMBERT model intercomparison project (MIP), providing both the motivations for the test cases as well as sufficient details to reproduce them, in the hopes that the results of this intercomparison can be used as a calibration for future GCMs. While tests of the protocol were run using the {\sc UM} and {\sc Exo-FMS} to gauge the viability of the protocol, the results from these tests are not presented here, with the intention being to present the results from all participating GCMs in one or more followup papers. The outline of this paper is as follows: an overview of the GCMs currently participating in the intercomparison and a discussion of the target planets are found is Section \ref{Sec:GCMs}. The protocol and associated test cases are described in Section \ref{Sec:Protocol}. The outputs and diagnostics are found in Section \ref{Sec:Outputs} and a final summary and discussion is found in Section \ref{Sec:Summary}.

\section{The Participating GCMs and the Target Planets}
\label{Sec:GCMs}

With the goal of the intercomparison being to understand the differences between GCMs, the progression of simulations can be divided up into two stages:  first, a simple investigation of the dynamics and the dynamical cores at the heart of each of the GCMs.  GCM simulations of GJ~1214b, for example, show up to three zonal jets, with differing amplitudes depending on the study \citep{menou2012,kataria2014,charnay2015,drummond2018,Mayne2019,wang2020}.  In understanding the origins of these differences, it is essential to be able to disentangle differences between dynamical cores and differences between radiative transfer schemes. Once this baseline understanding is established, we progress to a second stage where we compare models with radiative transfer.

\begin{deluxetable*}{lcc}
\tablecaption{Participating GCMs  \label{Tbl:GCMs}}
\tablehead{
\colhead{GCM} & \colhead{References} & \colhead{Point of Contact}
}
\startdata
       {\sc ExoCAM}  & \citet{wolf2022} & Eric T. Wolf \\
       {\sc Exo-FMS}  & \citet{Lee2021} & Elspeth K. H. Lee\\
       FMS PCM & \citet{ding2019,ding2020} & Feng Ding \\
       {\sc Generic PCM} & \citet{wordsworth2011,forgetinprep} & Benjamin Charnay \\
       {\sc MITgcm} & \citet{adcroft2004,showman2009,komacek2017} & Maria E. Steinrueck \\
       RM-GCM & \citet{rauscher2010,rauscher2012,malskyinprep} & Emily Rauscher \\
       THOR & \citet{mendonca2016,deitrick2020} & Russell Deitrick \\
       The {\sc Unified Model} (UM) & \citet{Mayne2014a,Mayne2019} & Duncan Christie \\
\enddata
\end{deluxetable*}

The GCMs currently participating, based on an expressed interest at the BUFFET workshop (\url{https://nexss.info/buffet-registration/}) and during the protocol development process, are listed in Table \ref{Tbl:GCMs}.  Although these are the participants at the time of the publication of this protocol, participation remains open and other teams are welcome to join.

\subsection{Choice of Targets}
\label{Sec:Targets}
As mini-Neptunes encompass a class of planets with a wide range of orbital and planetary parameters, we opt to focus on a warm, close-in case -- GJ~1214b -- and a cooler case -- K2-18b.  The specific target planets were selected based on the existence of previous GCM modelling efforts so as to reduce the barriers to participation as well as the existence of past and planned observations.  

%\subsubsection{GJ~1214b}

GJ~1214b is the first mini-Neptune discovered \citep{Charbonneau2009} and is the archetypal warm mini-Neptune.  Due to its proximity to its host star, it is expected to be tidally locked and possess a constant warm dayside and cool nightside, with the possibility of multiple zonal jets, as has been seen in GCM models by \citet{menou2012,kataria2014,charnay2015,drummond2018,Mayne2019,wang2020}. Questions about the applicability of the primitive equations in modelling these objects \citep{Mayne2019} as well as questions about the exact flow structure and the convergence timescale \citep{wang2020} further motivate a mini-Neptune intercomparison for GJ~1214b.  In addition, a JWST MIRI/LRS phase curve as well as multiple JWST NIRCam transit observations are planned \citep{beanjwst2021,greenejwst2017}.

%\subsubsection{K2-18b}
The more recently discovered K2-18b \citep{montet2015} is a temperate mini-Neptune in the habitable zone of its host star.   Its place in the habitable zone combined with a possible detection of water vapour in its atmosphere \citep{tsiaras2019,benneke2019} makes it a tempting target for both characterisation and observation, although the interpretation of the $1.4\, \mathrm{\mu m}$ signal has been disputed \citep{bezard2020,barclay2021}.  Thus far, it has been modelled in 3D by \citet{charnay2021} and  \citet{innes2021}, and there are planned JWST transit observations using NIRSpec  \citep{hujwst2021,madhusudhan2021}, MIRI \citep{madhusudhan2021}, and NIRISS \citep{madhusudhan2021}. While it is unclear whether or not K2-18b is tidally locked \citep[see e.g.,][]{leconte2015,charnay2021}, we assume that it is for simplicity.

\begin{deluxetable*}{lcc}
\tabletypesize{\scriptsize}
\tablewidth{0pt}
\tablecaption{Input Parameters\label{Tbl:Params}}
\tablehead{ 
\colhead{} & \colhead{GJ~1214b}  & \colhead{K2-18b}
} 
\startdata
{\em Common Planetary Parameters}\\
%Mass ($\mathrm{M_\oplus}$) & $8.17$& $8.63$\\
%Radius ($\mathrm{R_\oplus}$) & $2.68$& $2.61$ \\
Mass ($\mathrm{kg}$) & $4.88\times 10^{25}$& $5.15\times 10^{25}$\\
Radius ($\mathrm{m}$) & $1.75\times 10^7$ & $1.66\times 10^7$ \\
Orbital Radius (AU) & 0.01485 & $0.159$ \\ % State exact value of AU (m) in caption
Orbital and Rotation Period (days) & 1.58 & 32.94\\
Gravity ($\mathrm{m\,s^{-2}}$) & 10.7 & 12.4 \\
& & \\
{\em Common Stellar Parameters}\\
%Mass ($M_\odot$) & $0.15$ & $0.4951$\\
%Radius ($R_\odot$) & $0.216$ & $0.4445$\\
Mass ($\mathrm{kg}$) & $2.98\times 10^{29}$ & $7.14\times 10^{29}$\\
Radius ($\mathrm{m}$) & $1.50\times 10^8$ & $3.09\times 10^8$\\
$T_\mathrm{eff}$ (K) & $3250$ & $3457$ \\
Metallicity [Fe/H] & $0.29$ & $0.123$\\ 
log(g) & $5.026$ &  $4.8$\\
\\
{\em Case 1}\\
$\Delta T_\mathrm{eq,max}$ (K) & 600 & 50\\
Specific gas constant $R$  ($\mathrm{J\,kg^{-1}\,K^{-1}}$) & $3.513\times10^3$ & $1.732\times 10^3$\\
Specific heat capacity $c_P$ ($\mathrm{J\,kg^{-1}\,K^{-1}}$) & $1.200\times 10^4$ & $6.682\times 10^3$\\
Mean molar mass $\mu$ ($\mathrm{g}\,\mathrm{mol}^{-1}$) & $2.367$ & $4.801$\\ 
\\
{\em Case 2}\\
Shortwave Absorption $\kappa_\mathrm{sw}$ ($\mathrm{m^2\,kg^{-1}}$) & $1\times 10^{-4}$& $2\times 10^{-5}$\\
Longwave Absorption $\kappa_\mathrm{lw}$ ($\mathrm{m^2\,kg^{-1}}$) & $3\times 10^{-3}$& $1.4\times 10^{-2}$\\
Instellation ($\mathrm{W\, m^{-2}}$) & $2.17\times 10^4$ & $1.37\times 10^3$ \\
$T_\mathrm{int}$ (K) & $100$ & $90$ \\
\\
{\em Case 3}\\
Stellar Spectrum & 3000 K BT-Settl  & 3500 K BT-Settl\\
& with [Fe/H]=0.3 & with [Fe/H] = 0.0\\
& and log(g) = 5 & and log(g) = 5 \\
\enddata
\tablecomments{Planetary parameters for GJ~1214b are from \citet{cloutier2021} and parameters for K2-18b are from \citet{benneke2019} and \citet{cloutier2019}.
Stellar parameters are for GJ~1214 and K2-18 are taken from \citet{cloutier2021} and \citet{benneke2017}, respectively. To avoid ambiguities in the mass and radius values used in the intercomparison, the input values are quoted in units of $\mathrm{kg}$ and $\mathrm{m}$, respectively, with the conversion being done using the appropriate conversion factors in \citet{prsa2016}.}
\end{deluxetable*}

\section{Protocol}\label{Sec:Protocol}
In this section we outline the simulations associated with the intercomparison and their motivations.  In general, we seek to maintain consistency in parameters throughout the protocol to facilitate comparison as complexity is increased.  While the various cases were initially envisioned as a progression, we encourage participants to join the cases their GCMs are capable of completing regardless of whether or not their GCMs are capable of completing cases earlier in the series.

For all models, we adopt a lower boundary pressure of $3\times 10^6\,\mathrm{Pa}$.  While previous studies have studied higher pressures -- \citet{charnay2015} and \citet{Mayne2019} used $8\times 10^6\,\,\mathrm{Pa}$ and $2\times 10^7\,\,\mathrm{Pa}$, respectively -- to simplify the test cases and limit differences between the primitive and less simplified equations of the dynamics for mini-Neptunes \citep{Mayne2019}, we limit ourselves to $3\times 10^6\,\mathrm{Pa}$.  We adopt an upper boundary pressure $p \le 10\,\mathrm{Pa}$ for codes which employ a pressure-based vertical grid.  For GCMs which use a height-based vertical grid (e.g., the UM), we require a domain height sufficient to include pressures of $10\,\mathrm{Pa}$ throughout the simulations; however, this may require varying the domain height on a per-case basis. This will also result in those GCMs using height-based grids having potentially significantly lower pressures on the nightside upper boundary, which may necessitate modifications to ensure stability.   Following the lead of the THAI project, we do not place specific requirements on the timestep or grid spacing, instead encouraging participants to adopt parameters they would commonly use for exoplanet studies as requirements for stability may differ between GCMs.  

For simplicity, we assume friction-free, impermeable boundaries to avoid complicating the tests with boundary-layer friction or mass-exchange, and heat exchange is limited to a fixed internal heat flux with effective temperature $T_\mathrm{int}$.  We do not, however, exclude forms of dissipation that may be required for numerical stability (e.g., sponge layers, artificial viscosity).

 As mini-Neptunes may have enhanced atmospheric metallicities relative to solar \citep{fortney2013}, we adopt a value of $100\times$ solar for K2-18b and use parameters and profiles consistent with this throughout the intercomparison.  While there may be similar motivation to use this value for GJ~1214b, due to the number of previous simulations using solar metallicity and the differences between simulations already shown in the literature, specifically in the number and speed of the zonal jets,  we adopt the solar value as it allows us to probe a part of the parameter space where we know differences between GCMs already occur.

 Simulations are run for a fixed number of Earth days instead of specifying a convergence condition.  The chosen simulation lengths do not ensure convergence in simulations which include the deep atmosphere, this may not be sufficient time for the deep atmosphere to have converged to a steady state \citep{Mayne2017}; however, as we are limited to regions with $p \le 3\times 10^6\,\mathrm{Pa}$ we should not be significantly impacted.  While \citet{wang2020} have found in their simulations of GJ~1214b that over integration times of 50,000 Earth days or more their model atmospheres transition from two off-equatorial jets to a single equatorial jet, including such long integration times in an intercomparison would likely prove computationally prohibitive and limit participation.  We instead focus initially on comparing GCMs over shorter timescales, with hopes of extending the work in the future to look at these longer timescales.  For cases 1 and 3, we run the simulations for 4,000 Earth days.  For case 2, however, we run each simulation for 10,000 Earth days as the dual grey case is the case investigated by \citet{wang2020}.  Although this does not approach the long integration times of \citet{wang2020}, an integration time of 10,000 Earth days may be sufficient to understand differences between GCMs as \citet{menou2012} observed the formation of the central equatorial jet in their 7800 Earth day simulation.  It may be inevitable, however, that an understanding of a possible delayed formation of an equatorial jet may have to wait for a follow-up study with fewer participants investigating longer simulation times.

Although mini-Neptunes have the potential for clouds, and GJ~1214b in particular has been shown to have strong signs of clouds or hazes \citep{kreidberg2014}, we do not include a cloudy benchmark as a part of the protocol as currently there are an insufficient number of GCMs capable of participating.  We do hope that followup studies and intercomparisons will be able to include a cloud component as clouds will undoubtedly represent an important constituent of future mini-Neptune models.

\subsection{Initial Conditions}
\label{Sec:IC}
\begin{figure}
\plotone{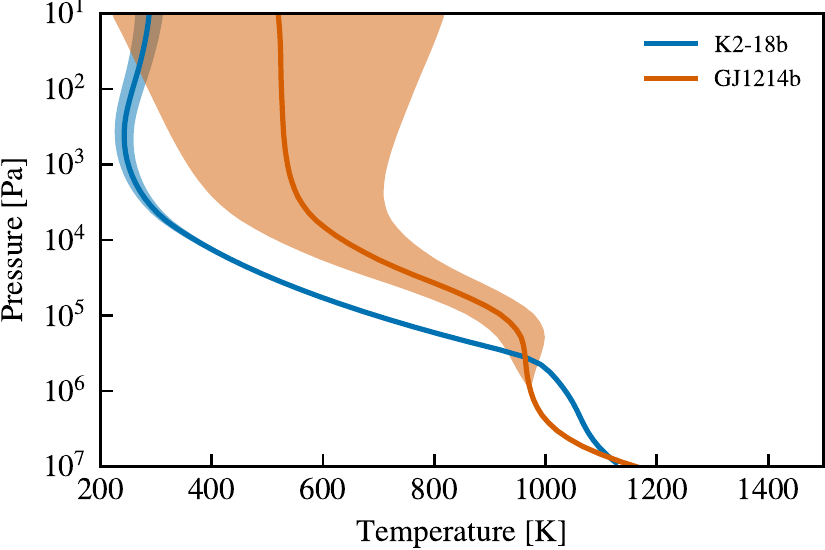}
    \caption{Initial temperature profiles for K2-18b and GJ~1214b.  The solid lines indicate the initial temperature profiles for each planet while the shaded regions indicate the range of equilibrium temperatures $T_\mathrm{eq}$ in the temperature-forced case (see Equation \ref{Eqn:Teq}). \label{Fig:Profiles}}
\end{figure}

To initialize our simulations, we use one-dimensional pressure-temperature profiles  (see Figure \ref{Fig:Profiles}) with no initial winds. For K2-18b, we use a profile generated using Exo-REM from \citet{charnay2021} and for GJ~1214b we use a profile generated using ATMO from \citet{drummond2018}. For all cases, we do not include any initial latitudinal or longitudinal variation.   Each of these profiles, along with the profiles for the chemical abundances needed for Case 3, is publicly available as an ASCII text file in the CAMEMBERT repository (see Section \ref{Sec:Repo}).

\subsection{Case 1: Temperature Forcing}

The first case investigated as a part of the intercomparison is similar to the temperature-forced benchmark of \citet{held1994}.  The motivation is to compare the robustness of the dynamical cores without also comparing differing radiative transfer schemes.   In this case, we opt to use a Newtonian cooling prescription where the temperature $T$ is forced to an equilibrium temperature profile $T_\mathrm{eq}$ on a given radiative timescale $\tau_\mathrm{rad}$,

\begin{equation}
    \frac{dT}{dt} = \frac{T_\mathrm{eq} - T}{\tau_\mathrm{rad}}\,\,,
\end{equation}

\noindent where $t$ is time. We adopt a radiative timescale $\tau_\mathrm{rad}$ for a hydrogen-dominated atmosphere \citep{zhang2017}

\begin{equation}
    \tau_\mathrm{rad,H_2}(p) = \begin{cases}
    10^4\,\,\mathrm{s} & p \le 10^2\,\mathrm{Pa} \\
    10^{5/2}p^{3/4}\, \,\mathrm{s} & 10^2\, \mathrm{Pa} < p < 10^6\,\mathrm{Pa} \\
    10^7 \,\mathrm{s} & p \ge 10^6 \, \mathrm{Pa} \\
    \end{cases}\,\, ,
\end{equation}

\begin{equation}
    \tau_\mathrm{rad}(p) = \tau_\mathrm{rad,H_2}(p)\left(\frac{c_p}{7R/2}\right)\frac{2}{\mu}\, ,
\end{equation}

\noindent with the specific values of $R$, $c_p$ and $\mu$ found in Table \ref{Tbl:Params}.

The equilibrium three-dimensional temperature profile $T_\mathrm{eq}$ is generated from the initial temperature profile $T_0(p)$ and temperature difference $\Delta T_\mathrm{eq}(p)$ intended to mimic qualitatively the day-night temperature contrast expected from tidally-locked planet,

\begin{equation}
    T_\mathrm{eq} = \begin{cases} 
    T_0(p) + \Delta T_\mathrm{eq}(p)\left(|\cos \phi \cos \lambda| - \frac{1}{2}\right) & \text{dayside} \\
    T_0(p) - \frac{1}{2}\Delta T_\mathrm{eq}(p) & \text{nightside}
    \end{cases}\, ,
    \label{Eqn:Teq}
\end{equation}

\noindent where the longitude is $\lambda$ and the latitude is $\phi$. The temperature difference $\Delta T_\mathrm{eq}$ is taken to be $\Delta T_\mathrm{eq,max}$ at $p \le 10 \,\mathrm{Pa}$ and to decrease linearly with $\log p$ until $p = 10^6\,\mathrm{Pa}$ where is becomes zero.  The range of equilibrium temperatures at a given pressure is shown in Figure \ref{Fig:Profiles}.  For GJ~1214b, the contrast parameter $\Delta T_\mathrm{eq,max}$ is chosen to maintain consistency with the temperature forcing tests of \citet{Mayne2019}.  As published temperature forcing tests do not exist for K2-18b, we instead look to \citet{charnay2021} which shows a more modest $\sim 50\,\mathrm{K}$ temperature contrast at the top of the atmosphere for their $100\times$ solar metallicity atmosphere, motivating the chosen value of $\Delta T_\mathrm{eq,max}$.

We note that the temperature forced case of K2-18b has presented significant difficulty in terms of numerical stability for the UM and Exo-FMS in tests of the protocol.   Rather than remove it, we retain it as a part of the protocol with a note of caution to participants.

\subsection{Case 2: Grey Radiative Transfer}

For the initial investigation of the impact of radiative transfer, we employ a dual band approximation with the shortwave and longwave absorption coefficients given in Table \ref{Tbl:Params} to compute heating rates, with the previous forcing scheme no longer included. The values of $\kappa_\mathrm{sw}$ and $\kappa_\mathrm{lw}$ have been calculated by fitting the initial profiles to the analytic profiles in \citet{guillot2010}\footnote{This method results in slightly different values of $\kappa_\mathrm{sw}$ and $\kappa_\mathrm{lw}$ for GJ~1214b compared to those used previously in \citet{menou2012}; however, for consistency in methodology with what is done for K2-18b, we choose to compute our own values instead of using the values in \citet{menou2012}. }.  As participating GCMs may offer different methods to attenuate the incoming stellar irradiation, simulations in Cases 2 and 3 are to be run using the plane-parallel approximation.  We adopt this intermediate step before transitioning to non-grey radiative transfer as previous studies show that disagreements between GCMs may already exist at this point \citep[e.g.,][]{menou2012,wang2020}.

\subsection{Case 3: Non-Grey Radiative Transfer with Fixed Abundance Profiles}

\begin{figure*}
    \plotone{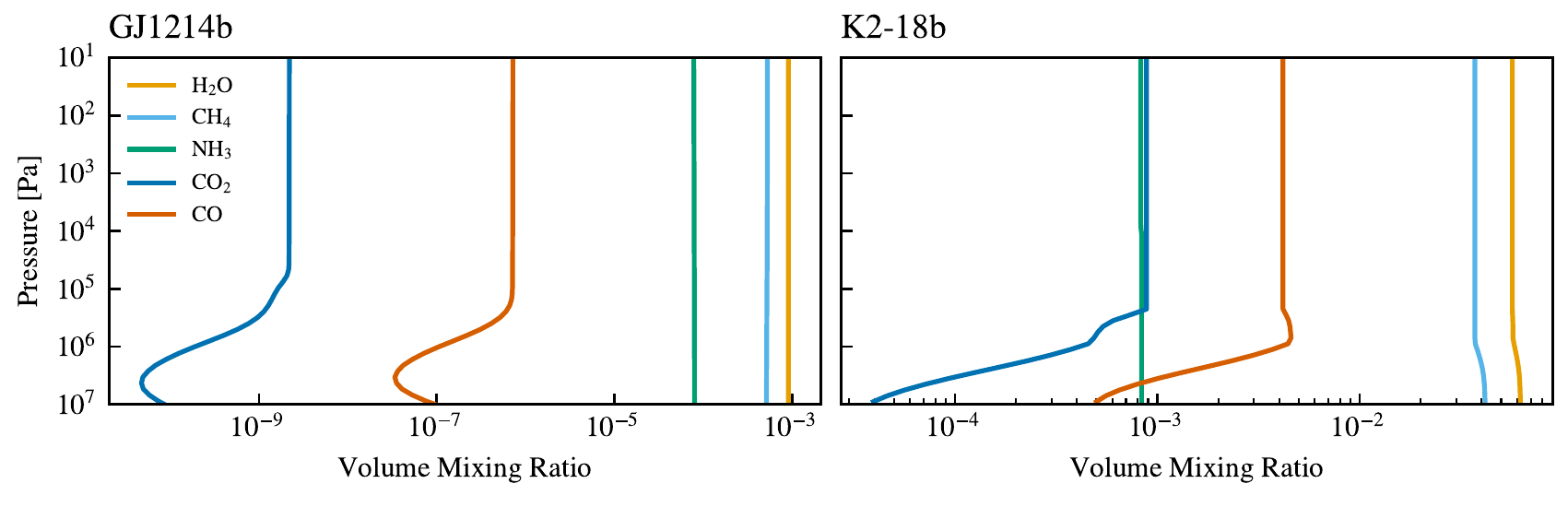}
    \caption{Chemical abundance profiles for K2-18b and GJ~1214b\label{Fig:ChemProfiles} to be used in Case 3.  These are publicly available in the CAMEMBERT repository (see Section \ref{Sec:Repo}).}
\end{figure*}

To model atmospheric chemistry, we limit ourselves to \ce{H_2}/\ce{He}-dominated atmospheres with \ce{H_2O}, \ce{CH_4}, \ce{NH_3}, \ce{CO} and \ce{CO_2} as well as \ce{H_2} and \ce{He} collisionally-induced absorption (CIA) and Rayleigh scattering as opacity sources.  Volume mixing ratios for each species as a function of gas pressure, taken from the same simulations that generated the initial conditions, are shown in Figure \ref{Fig:ChemProfiles} and are provided as a part of the initial conditions archive (see Section \ref{Sec:IC}).  All participating GCMs are to use these abundance profiles, as this allows GCMs without coupled chemistry solvers to participate.   

For the stellar spectra, we use the model stellar spectra from PHOENIX BT-Settl \citep{allard2012} which closest matches the target star (see Table \ref{Tbl:Params}).  These spectra are made available in the CUISINES repository along with the other required inputs.  The specific linelists, calculation method, and spectral resolution are left to the individual groups.

\section{Outputs, Diagnostics, and Archiving}\label{Sec:Outputs}

In this section we outline the procedures for formatting and archiving data.  The goal is to standardise the output and storage of data as much as possible to facilitate not only the initial analysis but also future analysis by third parties.

\subsection{Outputting and Formatting of Data}
\label{Sec:OutputFormat}
In order to provide sufficient data for comparison, we output diagnostic fields with a 1000 Earth day frequency without any averaging applied to track the evolution of the simulated atmospheres. For the final 1000 Earth days, outputs should be every 50 Earth days for the purposes of averaging.   For the comparison of atmospheric dynamics, we require pressure and temperature as well as eastward (u), northward (v), and vertical (w) velocity fields for all cases.   For cases 2 and 3, we additionally require longwave and shortwave heating rates as well as the top of atmosphere (TOA) outgoing longwave radiation (OLR),  outgoing shortwave radiation (OSR), and incoming shortwave radiation (ISR).  As the participating GCMs have different dissipation schemes which may lead to differing results, dissipation rates for all drag/damping/sponge schemes should be output as well.   The required outputs are summarised in Table \ref{Tbl:Out}.

To facilitate the sharing of data and subsequent analysis, we ask that all GCM outputs be in the netCDF format with data stored in SI units.  The metadata associated with each variable should include a description of the variable and the associated units, following the netCDF Climate and Forecasting Metadata Conventions\footnote{\href{https://cfconventions.org/}{https://cfconventions.org/} (last access: 2022 June 21)}. In the case of vector quantities, the sign conventions should be such that northward, eastward, and upward all constitute positive directions for their respective components. Similarly, data should be stored such that increasing grid indices corresponds to a positive change in direction. Fields should be placed on a single rectangular latitude-longitude grid, with longitudes beginning at $\lambda=0\degr$ and the poles located at latitudes  $\phi=\pm 90\degr$.  As the protocol considers only tidally-locked cases, the anti-stellar point should be located at $(\lambda,\phi)=(0\degr,0\degr)$.  Data along cyclic coordinates should not appear more than once within the dataset.  Along the vertical coordinate, a single pressure, potential temperature, or height grid should be adopted.  

As a number of the participating GCMs are capable of outputting scalar diagnostics with a higher frequency than other outputs, we ask that those GCMs with this capability provide the total axial angular momentum, kinetic energy, and maximum values of each velocity component with an output frequency they would normally adopt.  These data can be provided in a text file at the time of submission.  We do not make these data a requirement of the protocol as it would require additional development by a significant number of participants.  Instead, for those GCMs that do not provide these data separately, these scalar diagnostics will be derived via post-processing of the provided outputs.

\begin{deluxetable*}{lcc}
\tabletypesize{\scriptsize}
\tablewidth{0pt}
\tablecaption{Summary of Required Outputs and Diagnostics\label{Tbl:Out}}
\tablehead{
\colhead{Type of Output} & \colhead{Outputs} & \colhead{Dimensionality}
}
\startdata
    {\em Common Outputs} \\
       Atmospheric Profiles & Temperature, Pressure & 3D\\
        & u, v, w Velocity Fields & 3D\\
    Dissipation & Dissipation Rates for any Drag/Damping/Sponge Scheme & 3D\\
    \\
    {\em Case 1 Only} \\
    Temperature Forcing & Heating Rate (in $\mathrm{K\,s^{-1}}$) & 3D\\
    \\
    {\em Cases 2 and 3 Only} \\
        Radiation & OLR, OSR, and ISR & 2D\\
        & Shortwave and Longwave Heating Rates (in $\mathrm{K\,s^{-1}}$) & 3D\\
\enddata
\tablecomments{All fields are to be output every 1000 Earth days with the output frequency increased to every 50 Earth days during the final 1000 days of each simulation. Details can be found in Section \ref{Sec:OutputFormat}.}
\end{deluxetable*}

\subsection{The CAMEMBERT Repositories}
\label{Sec:Repo}
The data resulting from the simulations and analysis will be uploaded to the CAMEMBERT permanent repository at \url{https://ckan.emac.gsfc.nasa.gov/organization/cuisines-camembert} by participating scientists.  These data will be made available for public access upon the publication of the results.  Pre-publication access can be requested by contacting the authors.   Inputs described in this protocol and scripts related to the analysis of data and production of plots for the publications will be made available on the CAMEMBERT GitHub repository at \url{https://github.com/projectcuisines/camembert}.  Inputs will be available immediately while scripts to reproduce results will be made publicly available upon the publication of the results.

\subsection{Simulated Observables}

As discussed in Section \ref{Sec:Targets}, GJ~1214b and K2-18b were chosen in part because there are planned observations with JWST in Cycle~1 \citep{greenejwst2017,beanjwst2021,hujwst2021,madhusudhan2021}.  The permanent repository will host the results of the analysis of the results and the post-processed synthetic observations. Consistent with the other intercomparisons that are a part of CUISINES, we will use the Planetary Spectrum Generator \citep[PSG;][]{villanueva2018,villanueva2022} to simulate JWST spectra for instruments and modes used in Cycle 1 observations of GJ~1214b and K2-18b, and subsequent cycles when available, using the atmospheric outputs provided by each GCM for each of the three cases. 

%\subsection{Simulated Observables}
%The broader objectives of CUISINES are twofold. First, it provides a meta-framework to quantify, and potentially mitigate, differences between exoplanet model outputs. Second, it aims to assess how these output differences affect the spectral simulations that are used to predict the detectability of atmospheric constituents or surface properties (and ultimately retrieving them with real data) with ground and/or space teslescopes.
%Here, we have selected GJ~1214-b and K2-18b as our CAMEMBERT  cases specifically because they are planed to be observed with JWST from Cycle~1. GJ~1214b will have a full phase curve observed with MIRI low resolution spectroscopy (LRS) as  (Guest observer (GO) program 1803 by PI J. Bean) and two transit observations with NIRCam Grism Time Series (guaranteed time observations (GTO) proposal 1185, PI T. Greene). K2-18b will get 6 transit observation with NIRSpec (GO proposal 2372 by PI R. Hu) using the G235H and G395H gratings to cover the wavelength range 1.7--5.2~$\mu m$. \\
%Consistently with the other CUISINES MIP of exoplanet atmospheric models, we will use the Planetary Spectrum Generator (PSG, \cite{Villanueva2018,Villanueva2022} to simulate JWST spectra following the above instruments and observation modes for Cycle 1 (and following cycles when available) using the atmospheric outputs provided by each GCM for the three CAMEMBERT cases. Differences between the spectra will be discussed and compared to those simulated from other CUISINES project, such as THAI.

\subsection{Environmental Impact}

While not related to the accuracy of the GCMs, we add as a part of the protocol the requirement that participating scientists include estimates of power consumption and \ce{CO_2} emissions associated with each production run included in the intercomparison.   We include this requirement not as point of comparison between GCMs, as the environmental impact will primarily depend on the methods of energy generation in the local power grids. Instead, we include this to highlight the environmental impact of supercomputing and to encourage providers of supercomputing resources to transition to environmentally-sustainable energy sources.  These data will be reported in the first results paper.

\section{Summary}\label{Sec:Summary}

In this paper, we have presented the protocol for the CAMEMBERT project which seeks to compare GCMs used by the exoplanetary science community, with models of mini-Neptunes being the primary focus.     Two benchmarks were chosen -- the warm GJ~1214b and the relatively cooler K2-18b -- based on the volume of prior modelling work and observational potential, and a series of simulations of increasing complexity are described to calibrate and compare the participating GCMs, with all of the requisite parameters provided here and in the CUISINES repository.  Membership in CAMEMBERT remains open, and other groups interested in participating are invited to contact the authors. Collaboration meetings will be held in 2022 and beyond,  and  the aim is to present the results of the intercomparison in one or more follow-up papers.  It is hoped that the results from this intercomparison will provide a strong foundation for follow-up studies exploring elements not including in this initial protocol such as chemistry, clouds, and convergence timescales with any model differences due to these elements being more easily isolated and interpreted.  As CUISINES brings together researchers employing a diverse range of tools and approaches, we anticipate that it will act to create new collaborations and stimulate progress in understanding exoplanets. For CAMEMBERT specifically, connections with MALBEC (Modeling Atmospheric Lines By the Exoplanet Community, an intercomparison of radiative transfer codes) and PIE (Photochemical model Intercomparison for Exoplanet science, an intercomparison of 1D photochemistry codes) will begin in 2022 with the goal of comparing results and better informing our own models with the insights from the other projects.

%, exchanging data, and hopefully improving our understanding of mini-Neptunes.

%as simulation outputs from CAMEMBERT are a valuable source of atmospheric pressure-temperature profiles for those projects.

%As CUISINES brings together researchers employing a diverse range of tools and approaches, 

%\begin{acknowledgments}
\section*{Acknowledgements}
CAMEMBERT belongs to the CUISINES meta-framework, a Nexus for Exoplanet System Science (NExSS) science working group.  The testing of the protocol was done using Met Office Software. The contributions of D. A. Christie, N. J. Mayne, D. Sergeev, and M. Zamyatina  made use of the ISCA High Performance Computing Service at the University of Exeter and the DiRAC Data Intensive service at Leicester, operated by the University of Leicester IT Services, which forms part of the STFC DiRAC HPC Facility (www.dirac.ac.uk). The equipment for DIRAC was funded by BEIS capital funding via STFC capital grants ST/K000373/1 and ST/R002363/1 and STFC DiRAC Operations grant ST/R001014/1. DiRAC is part of the National e-Infrastructure. The work of D. A. Christie, N. J. Mayne, D. Sergeev, and M. Zamyatina was also partly funded by the Leverhulme Trust through a research project grant [RPG-2020-82], a Science and Technology Facilities Council Consolidated Grant [ST/R000395/1] and a a UKRI Future Leaders Fellowship [grant number MR/T040866/1].  T. J. Fauchez acknowledges support from the GSFC Sellers Exoplanet Environments Collaboration (SEEC), which is funded in part by the NASA Planetary Science Divisions Internal Scientist Funding Model.  E. K. H. Lee and P. A. Noti are supported by the SNSF Ambizione Fellowship grant (\#193448).  H. Innes is supported by funding from the European Research Council (ERC) under the European Union’s Horizon 2020 research and innovation programme (Grant agreement No. 740963). The work of B. Charnay was granted access to the HPC resources of MesoPSL financed by the Region Ile de France and the project Equip@Meso (reference ANR-10-EQPX-29-01) of the programme Investissements d’Avenir supervised by the Agence Nationale pour la Recherche.
%\end{acknowledgments}

\bibliography{camembert}{}
\bibliographystyle{aasjournal}



\end{document}